\documentstyle[12pt]{article}

\def\be{\begin{eqnarray}} %-- equation for all cases
\def\ee{\end{eqnarray}}
\def\nn{\nonumber}

\newcommand{\rms}{\langle r^2 \rangle}

\ifx\undefined\varkappa\else\def\kappa{\varkappa}\fi
\ifx\undefined\lesssim % in not amssymb
\newcommand{\lesssim}{\mathrel{\raisebox{-.3ex}[1.ex][1.ex]{%
\mbox{$\,\stackrel{\scriptstyle<}{\scriptscriptstyle \sim}\,$}}}}
\newcommand{\gtrsim}{\mathrel{\raisebox{-.3ex}[1.ex][1.ex]{%
\mbox{$\,\stackrel{\scriptstyle>}{\scriptscriptstyle \sim}\,$}}}}
\fi

\begin{document}
\title{Phenomenological Three-Cluster Model of \6He%
\footnote[0]{The research described in this publication was made possible
in part by Grant No.~U48000 from the International Science Foundation. The
work was also supported in part by the Ukrainian State Committee for
Science and Technology}}
\author{G. F. Filippov,\ \ I. Yu. Rybkin,\ \ S. V. Korennov \\
{\normalsize Bogolyubov Institute for Theoretical Physics
 252143, Kiev-143,  Ukraine} }
\date{\normalsize(To appear in {\it Yad.\ Fiz.})}

%--\begin{titlepage}

\maketitle
\begin{abstract}
By using the method of hyperspherical functions within the appropriate
for this method $K_{\min}$ approximation, the simple three-cluster model
for description of the ground state and the continuous spectrum states
of \6He is developed. It is shown that many properties of \6He (its large rms
radius and large values of the matrix elements of electromagnetic transitions
from the ground state into the continuous spectrum) follow from the fact that
the potential energy of \6He system decreases very slowly (as $\rho^{-3}$) and
the binding energy is small.
\end{abstract}
%--\end{titlepage}

%------------------------------------------------
%  SECTION 1
%------------------------------------------------
\section{Introduction}

The \6He nucleus is an example of a three-cluster system whose lowest
threshold ($\alpha+n+n$) is a three-particle one. Such systems have a number
of remarkable properties determined mainly by two factors---the Pauli
principle and the character of the potential energy dependence due to the
three-body structure.

The Pauli principle imposes essential restrictions on the wave function of a
system allowing it to contain only the components antisymmetric with respect
to nucleon permutation. Thus, if we are using the expansion over the
harmonic-oscillator basis, we must solve the problem of excluding the states
forbidden by the Pauli principle \cite{ho-basis}.  This, in particular, leads
to the fact that the simplest wave function of the $0^+$-state of \6He
obtained with the translation-invariant shell model (it coincides with the
lowest oscillator-basis function) is a superposition of states with
three-particle hypermomentum $K=0$ ($\sim5$\%) and $K=2$ ($\sim95\%$).  This
function generates an infinite set of states differing by a number of
oscillator quanta and having almost the same ratio of $K=0$ and $K=2$
components. The approximation using only this set of functions can be
appropriately called the $K_{\min}$-approximation. Other basis states contain
components with $K>2$ whose weight is small, as shown in \cite{k-harm}.

Notice that for $2^+$-states, wave functions of the $K_{\min}$-approximation
contain only $K=2$ components, while for the $1^-$-states they are the
superpositions of $K=1$ (${\gtrsim 25\%}$) and $K=3$ (${\lesssim 75\%}$)
components.

The second characteristic feature of systems under consideration is the slow
decrease of their potential energy for large values of the hyperradius $\rho$.
As known (see for example \cite{rho-cube}), in three-body systems, the
potential energy has the asymptotical behaviour ${\rm const}/\rho^3$ as $\rho
\to \infty$, even if the two-body forces acting between any two of constituent
particles are short-range. This fact leads to two important results. First,
the boundary of a system is extremely loose and the rms radius is
significantly larger compared with that of those neighbouring nuclei whose
three-particle threshold is much higher than the binary one.  Second, the
phase shift of the three-body scattering rises sharply at low energies (it is
proportional to $\sqrt{E}$ as $E\to0$) which directly affects the behaviour of
the electromagnetic transitions matrix elements between the ground state of
\6He and its continuous spectrum.

The aim of the present paper is to give a detailed quantitative analysis of
general regularities in \6He structure, based on a simple model taking into
account the main features of a loosely-bound system with the lowest
three-particle threshold.

%------------------------------------------------
%  SECTION 2
%------------------------------------------------
\section{The Model}\indent

The allowed basis functions for states with zero angular momentum
$\varphi_{n}(L=0)$ of the $K_{\min}$-approximation have the following form
\be
\varphi_n(L=0) = A_0(n)\,\varphi_{n0}(\rho)\,u_0(\omega) +
A_2(n)\,\varphi_{n2}(\rho)\,u_2(\omega),
\ee
where the superposition coefficients are
\be
A_0(n) = \sqrt{4(n+1)\over 29n+104},~~
A_2(n) = \sqrt{25(n+4)\over 29n+104},
\ee
$u_K(\omega)$ are the hyperspherical functions,
\be
\varphi_{nK}(\rho) = N_{nk} \rho^K e^{-\rho^2/2} L^{K+2}_n(\rho^2)
\ee
are the normalized hyperradial harmonic-oscillator basis functions,
$L^{\alpha}_n(\rho^2)$ are the Laguerre polynomials,
$N_{nk}$ are the normalizing coefficients, $n = 0,1, \ldots$
Therefore, the wave function of the $L=0$ state in the
$K_{\min}$-approximation is reduced to the expansion
\be
\Psi(L=0) = \sum_{n=0}^{\infty} C_n \varphi_n(L=0).
\label{3}
\ee
For all $n$, the contribution of $K=2$ states is significantly greater than
that of the $K=0$ states
\be
0.86 < A_2^2(n) < 0.96;~~ A_0^2(n) = 1 - A_2^2(n),
\nn
\ee
hence, in a simple model, the contribution of $K=0$ states can be neglected.
Then, for the expansion (\ref{3}) the following relations hold
\be
\Psi(L=0) = \sum_{n=0}^{\infty} C_n \varphi_{n2}(\rho)\,u_2(\rho) =
\Phi(\rho)\, u_2(\Omega) = {\chi(\rho) \over \rho^{5/2}} u_2(\Omega).
\ee
The function
$\chi(\rho)$ should satisfy the one-dimensional Schr\"odinger equation
\be
{\hbar^2\over2m}\left[
-{d^2\chi\over d\rho^2} +
\left(K+{3\over2}\right)\left(K+{5\over2}\right){\chi\over \rho^2}
\right] + V(\rho)\,\chi = E\chi,
\label{5}
\ee
where $m$ is the nucleon mass, $K$ is the hypermomentum ($K=2$ for $L=0$),
$V(\rho)$ is the effective three-body potential.

The equation (\ref{5}) is a starting point for the subsequent discussion.
It was solved by a numerical integration over $\rho$ from zero to a certain
sufficiently large cutoff radius $\rho_{\max}$.
The effective potential is modelled by a function
\be
 V(\rho) = {V_0 \over {\displaystyle 1 + \left({\rho\over a}\right)^3}},
 \label{6}
\ee
having correct asymptotics and without a singularity at $\rho=0$. To reproduce
the states with zero angular momentum, we have chosen the parameters $V_0$ and
$a$ such that the equation (\ref{5}) would give experimental values for the
\6He binding energy and the rms radius, $E = -{\cal E} = -0.97$ MeV,
$\sqrt{\rms} = 2.57$~Fm. The appropriate values of the parameters are
\be
V_0 = -87~ {\rm MeV},~~ a = 3.073~ {\rm Fm}.
\ee
With these values we have calculated the wave function of the \6He ground
state and $0^+$-states of its continuous spectrum.

%------------------------------------------------
%  SECTION 3
%------------------------------------------------
\section{The Ground State of \6He}

The wave function of the ground state of \6He together with the effective
potential $V(\rho)$ are presented in Fig.~1. The horizontal line below the
$\rho$-axis corresponds to the ground state energy, its intersection with the
potential energy curve marks the classical turning point. The vertical line
separates the values of $\rho$ less than $\sqrt{\langle\rho^2\rangle} =
5.59$~Fm. The wave function falls as $\rho^{K+5/2}$ for small $\rho$,
due to the strong kinematical barrier, while for large $\rho$, the long-range
character of the potential $V(\rho)$ leads to the slow decrease of the wave
function and, respectively, to a significant diffuseness of the
boundary of a nuclear system.

%As can be seen from the comparison of $\chi(\rho)$ and $V(\rho)$
%curves, the nuclear system has a broad boundary while the wave function
%decrease very slowly spreading far beyond the turning point.

As known, loosely-bound binary systems with a short-range potential can be
rather well approximated outside the potential range by the exponential
function
\be
\chi(\rho) \simeq
\sqrt{2\kappa}\exp(-\kappa r),~~~~\kappa = \sqrt{{2m\over\hbar^2}{\cal E}\,},
\label{8}
\ee
where ${\cal E}$ is the bound state energy, $r$ is the radial variable of the
binary channel. The formal criterion of validity of such an approximation (the
zero-range approximation) is the smallness of a ratio of the potential range
and the cluster radii to the system radius $1/\sqrt{2\kappa}$ expressed in
terms of binding energy. It seems useful to compare the exact wave function
obtained after the solution of Eq.~(\ref{5}) with the approximation given by
Eq.~(\ref{8}) (for $\kappa \approx 0.22$ Fm$^{-1}$, dashed line in Fig.~1). It
can be seen that the nucleon system is much more loose than it is predicted by
the wave function (\ref{8}). The reason for that is not only that the function
$\chi(\rho)$ at small $\rho$ behaves as $\rho^{K+5/2}$ as $\rho\to 0$, but
first of all that the effective potential energy decreases slower than the
exponential, while the formula (\ref{8}) is obtained by supposing that the
potential is short-range.  As a result, the estimates of the square of
hyperradius mean value  $\sqrt{\langle\rho^2\rangle}$ based on (\ref{8}) are
about three times less than the experimental value.

Of course, for sufficiently large $\rho$ the function $\chi(\rho)$ has the
asymptotics
\be
\chi(\rho) \simeq C\sqrt{\kappa\rho}\,K_4(\kappa\rho)
\simeq C \exp(-\kappa\rho),
\label{10}
\ee
where
$K_4(\rho)$ is the Macdonald function, and the value of the coefficient $C$
(it is related to the so-called nuclear vertex constants, see \cite{vertex}) is
considerably less than that of $\sqrt{2\kappa}$
\be
C \approx 0.12~ {\rm Fm}^{-1/2} < \sqrt{2\kappa}
   \approx 0.66~ {\rm Fm}^{-1/2}.
\ee
This asymptotics is denoted by short-dashed line in Fig.~1.

%------------------------------------------------
%  SECTION 4
%------------------------------------------------
\section{0$^+$-States of the Continuous Spectrum}

Of the special interest are the solutions of Eq.~(\ref{5}) for the continuous
spectrum for relatively low over-threshold energies.  In particular, attention
must be paid to the question of the behaviour of the three-to-three
scattering phase $\delta$ as a function of energy in a potential field with
the asymptotical behaviour ${\rm const}/\!\rho^3$ and the powerful kinematic
barrier $(\hbar^2/2m)(63/4\rho^2)$ corresponding to $K=2$.  According to the
estimates \cite{LK-63,Landau}, at low energies, the phase shift, as is also
demonstrated by our calculations, is proportional to $k$ (or $\sqrt{E}$)
\be
\delta \simeq Ak + \ldots\:;~~~~ A = 3.95~{\rm Fm}.
\ee
The low-energy values of $\tan\delta$ calculated with different values of the
cutoff radius $\rho_{\max}$ are presented as functions of $k$ in Fig.~2.  As
seen, $\rho_{\max}$ should be rather large, at least 1000~Fm for
energies below 10 keV\@. The obtained value of $A$ gives a rather reasonable
prediction for the ground-state energy of \6He
\be
{\cal E}_{\rm approx.} = {\hbar^2\over2m}\,{1\over A^2} \approx 1.3~{\rm MeV},
{}~~~({\rm cf.}~{\cal E}_{\rm exp.} = 0.97~{\rm MeV}).
\ee

The three-body phase shift $\delta$ calculated with different $\rho_{\max}$ as
functions of energy $E$ in an interval up to 5 MeV are presented in Fig.~3. As
seen, for reliable calculations for medium-energy region ($\sim 1$~MeV)
$\rho_{\max}$ may be taken about 50~Fm and only for the lowest energies it
must be increased.  The phase shift rises steeply from zero to values
exceeding $90^{\circ}$ for $E\approx2.5~{\rm MeV}$ and then slowly goes down.
When the phase shift is near $90^{\circ}$ the first maximum of the wave
function moves closer to zero so that the matrix element of the isoscalar
transition from the ground state into the continuous spectrum increases (see
Fig.~4). This matrix element reaches its maximal value for $E\approx 1.3$~MeV
which could directly affect the electrodisintegration cross-section of \6He.

%------------------------------------------------
%  SECTION 5
%------------------------------------------------
\section{2$^+$-States of the Continuous Spectrum}

The $K_{\min}$-approximation satisfying the Pauli principle for the
$2^+$-states of \6He contain only the hyperspherical function with
$K=2$. Therefore, to obtain the wave functions $\chi(\rho)$ of $2^+$-states,
we again turn to the equation (\ref{5}) with $K=2$ but with the other
parameters of the potential $V(\rho)$.

The \6He nucleus has no $2^+$ bound state but it has a $2^+$-resonance at the
energy of 0.822$\pm$0.025 MeV with the width 0.113$\pm$0.020 MeV \cite{Ajz}.
The energy and the width of the resonance can be reproduced with the following
parameters of  potential:
\be
V_0 = -92~ {\rm MeV},~~ a = 2.834~{\rm Fm}.
\label{17}
\ee

The wave function $\chi(\rho)$ of the resonance state and the potential energy
with the parameters (\ref{17}) are presented in Fig.~5. For small $\rho$ the
resonance wave function behaves similarly to the wave function of the ground
state (dashed line in Fig.~5) but for larger $\rho$, in the asymptotical
region it, as it should be, oscillates.

The scattering phase shift in the $2^+$-state obtained by solving the equation
(\ref{5}) with the new potential parameters is presented in Fig.~6. At the
energy of about 1 MeV the phase shift has a typical behaviour for a resonance
region.

We have also calculated the matrix element of the operator of the isoscalar
$E2$ transition from the ground state of \6He to the $2^+$-states of its
continuous spectrum. The dependence of this matrix element upon the energy is
presented in Fig.~7.  The narrow peak observed for the energy about $0.8$~MeV
again demonstrates the small width of $2^+$ resonance and the large value of
the matrix element.

%------------------------------------------------
%  SECTION 6
%------------------------------------------------
\section{Conclusion}

The simple three-cluster model based on the phenomenological long-range
potential with the Pauli principle taken into account has allowed us to reveal
a number of regularities both for the weakly-coupled ground state of \6He and
for the states of its continuous spectrum for the relatively low (up to a few
MeV) energies. The large value of the rms radius in the ground
state can be explained by the considerable diffuseness of the wave function
caused by the slowly-decreasing effective potential. The asymptotic region,
where the wave function decreases exponentially begins only at the hyperradius
values greater than 15--20~Fm.

The phase shift of elastic scattering ($3\to 3$) is proportional to
$k$ or $\sqrt{E}$ at low energies which leads to the sharp maximum of the
matrix element of the isoscalar transition from the ground state to the
$0^+$-states of the continuous spectrum. Finally, our calculations predict a
considerable enhancing of the probability of the radiative capture of two
neutrons by the alpha-particle for the energy corresponding to the
$2^+$-resonance of \6He.

\medskip
We thank Dr.\ A. A. Korsheninnikov for stimulating discussions.

%---------------------------------------------------------------------
%        REFERENCES
%---------------------------------------------------------------------

\newpage
%---------------------------------------------------------------------
%        FIGURES
%---------------------------------------------------------------------
\unitlength 1in
%------------ FIG. 1 ---------------
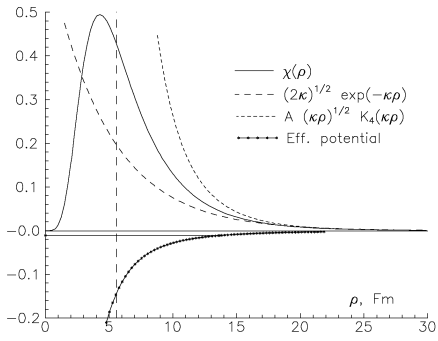
\begin{figure}
\begin{picture}(5,3.8)
\put(0,3.7){\special{em:graph fig1e.pcx}}
\end{picture}
\caption{The wave function of \6He ground state, its asymptotic behaviour
(Eqs.~(\protect\ref{8}) and (\protect\ref{10}))
and the effective potential. The ground-state energy is marked by a horizontal
line, the vertical line corresponds to
$\rho = \protect\sqrt{\langle\rho^2\rangle} = 5.59$ Fm).}
\end{figure}

%------------ FIG. 2 ---------------
\begin{figure}
\begin{picture}(5,3.8)
\put(0,3.7){\special{em:graph fig2e.pcx}}
\end{picture}
\caption{Low-energy behaviour of $\tan \delta$ as a function of $k$.
%%%Calculations for $\rho_{\max}=200$~Fm, 1000~Fm, 2000~Fm.
}
\end{figure}
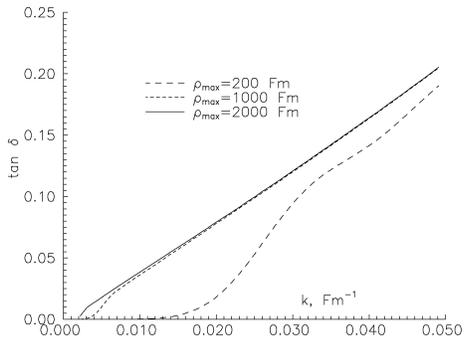

%------------ FIG. 3 ---------------
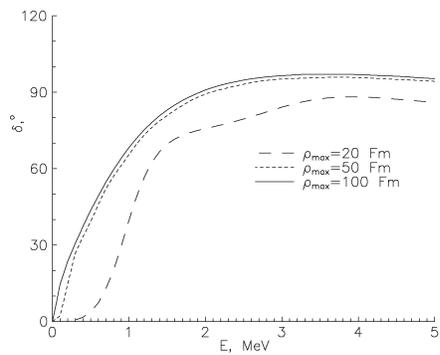
\begin{figure}
\begin{picture}(5,3.8)
\put(0,3.7){\special{em:graph fig3.pcx}}
\end{picture}
\caption{The scattering phase shift in the $0^+$-state.
%%%Calculations for $\rho_{\max}=20$~Fm, 50~Fm, 100~Fm.
}
\end{figure}

%------------ FIG. 4 ---------------
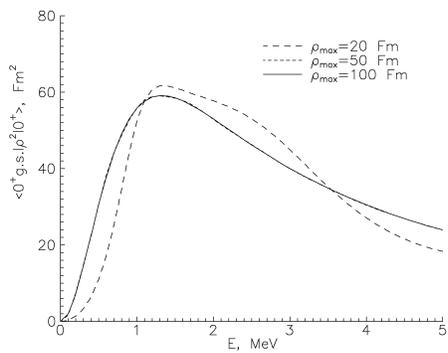
\begin{figure}
\begin{picture}(5,3.8)
\put(0,3.7){\special{em:graph fig4.pcx}}
\end{picture}
\caption{The matrix element of the isoscalar transition from the ground state
to the $0^+$ states of the continuous spectrum.
%%%Calculations for $\rho_{\max}=20$~Fm, 50~Fm, 100~Fm
(Results for $\rho_{\max} = 50$~Fm and 100~Fm are almost identical.)}
\end{figure}

%------------ FIG. 5 ---------------
\begin{figure}
\begin{picture}(5,3.8)
\put(0,3.7){\special{em:graph fig5e.pcx}}
\end{picture}
\caption{The wave function of the $2^+$ resonance and the ground state
(scaled) and the effective potential in the $2^+$ state}
\end{figure}
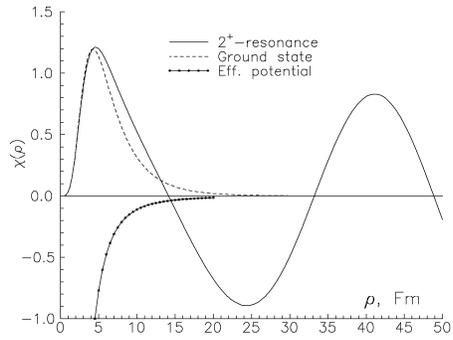

%------------ FIG. 6 ---------------
\begin{figure}
\begin{picture}(5,3.8)
\put(0,3.7){\special{em:graph fig6.pcx}}
\end{picture}
\caption{The scattering phase shift in the $2^+$-state.
%%%Calculations for $\rho_{\max}=20$~Fm, 50~Fm, 100~Fm
}
\end{figure}
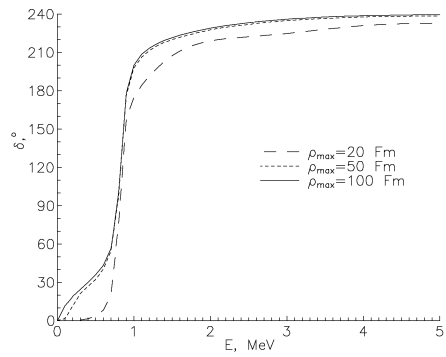

%------------ FIG. 7 ---------------
\unitlength 1in
\begin{figure}[t]
\begin{picture}(5,3.8)
\put(0,3.7){\special{em:graph fig7.pcx}}
\end{picture}
\caption{The matrix element of the isoscalar $E2$ transition from the ground
state
to the $2^+$ states of the continuous spectrum.
%%Calculations for $\rho_{\max}=20$~Fm, 50~Fm, 100~Fm
(Results for $\rho_{\max} = 50$~Fm and 100~Fm are almost identical.)}
\end{figure}
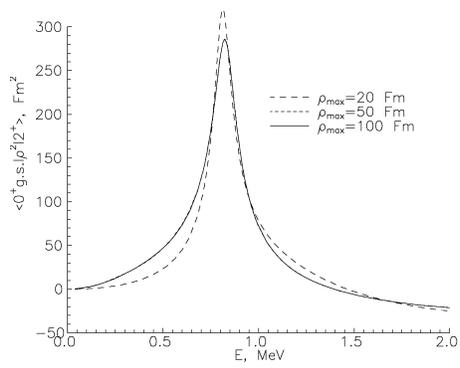
\vfill

\end{document}